\documentclass[%
 reprint,
 amsmath,amssymb,
 aps,
 prb,
]{revtex4-1}

\usepackage{graphicx}
\usepackage{dcolumn}
\usepackage{bm}
\usepackage{color}
\usepackage{times}
\usepackage{amsmath}
\usepackage[english]{babel}
\usepackage{float}
\usepackage{soulpos}

\newcommand{\beginsupplement}{%
	\setcounter{table}{0}
	\renewcommand{\thetable}{S\arabic{table}}%
	\setcounter{figure}{0}
	\renewcommand{\thefigure}{A\arabic{figure}}%
}
 
\begin{document}
\title{Re-programmable self-assembly of magnetic lattices in two-dimensions}

\author{Audrey A. Watkins and Osama R. Bilal} 
\affiliation{School of Mechanical, Aerospace, and Manufacturing Engineering, University of Connecticut, Storrs, USA}

\date{\today}

\begin{abstract}


Simple local interactions can cause primitive building blocks to self-assemble into complex and functional patterns.  However, even for a small number of blocks, there exist a vast number of possible configurations that are plausible, stable, and with varying degree of order. The ability to dynamically shift between multi-stable patterns (i.e., reprogram the self-assembly) entails navigating an intractable search space, which remains a challenge. In this paper, we engineer the self-assembly of macroscopic magnetic particles to create metamaterials with dynamically reversible emergent phases. We utilize a boundary composed of magnetic hinges to confine free-floating magnetic disks into different stable assemblies.  We exploit the non-destructive nature of the magnetic boundaries to create re-programmable two-dimensional metamaterials that morphs from crystalline to quasi-crystalline to disordered assembly using the same number of disks and boundary.  Furthermore, we explore their utility to control the propagation of sound waves in an effectively undamped media with rich nonlinearities. Our findings can expand the metamaterials horizon into functional and tunable devices.

\end{abstract}

\maketitle

Self-assembled structures are ubiquitous in both natural and artificial systems \cite{whitesides2002self, ball2001self, ptitsyn1995molten}. The crystal structures of ribosomes \cite{ban2000complete}, microstructures of various materials \cite{clark2001self, terfort1997three, terfort1998self, gracias2000forming}, and the formation of self-assembled molecular structures on graphene \cite{macleod2014molecular, lauffer2008molecular, wang2009room} all rely on particles aligning themselves into a desirable pattern. Self-assembly of structures is a result of a system attempting to minimize its own potential energy through the realignment of its constitutive parts.  The self-assembly processes can produce intricate and operational structures through localized interactions among numerous simpler constituents. Engineering  (i.e., programming) the self-assembly of basic building blocks not only unveils essential physics but also enables the exploration of exciting applications in manufacturing, drug delivery, and wave manipulation \cite{stebe2009oriented, mastrangeli2009self, caleap2014acoustically, malassis2014topological, yigit2019programmable, huang2020dynamical, sitti2021physical, han2020reconfigurable, ceron2023programmable, nagaoka2023quasicrystalline, watkins2020demultiplexing, watkins2021exploiting, watkins2022harnessing, eichelberg2022metamaterials}. Further, the ability to harness reversible (i.e., reprogrammable) self-assembly can offer novel avenues for tuning structural and material properties on demand. However, the transition between emerging patterns with various levels of order and disorder, represents a long-standing problem \cite{grzybowski2002dynamics, dotera2014mosaic, ou2020kinetic, culha2020statistical,  jeon2021reversible, yin2021transition, zhang2022guiding}.

\begin{figure}[b]
\includegraphics[scale=0.6]{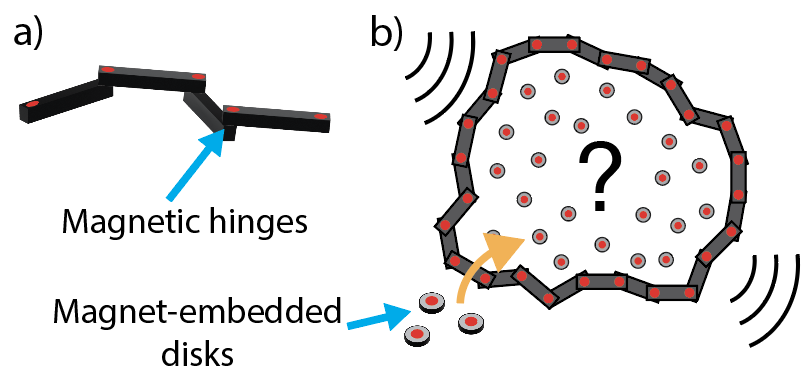}
\caption{\label{fig:concept} \textbf{Concept.} a) Schematic of a portion (4 links) of flexible, bike-chain-like magnetic boundary. b) Wave transmission through the self-assembled lattice comprised of magnet-embedded disks confined within the flexible boundary.}
\end{figure}

The self assembly of floating magnets is a problem that is more than a century old \cite{mayer1878floating}. Such basic elements represent an excellent platform to explore emergent out-of-equilibrium structures, complex collective dynamics, and design rules for the next generation of materials\cite{han2020reconfigurable}.  In this paper, we engineer the self-assembly of macroscopic magnetic particles to create metamaterials with dynamically reversible emergent phases. We study both the static and the dynamic characteristics of the resulting assemblies. The metamaterials are composed of free-floating disks (i.e., meta-atoms) with embedded permanent magnets that are confined in-plane within a programmable magnetic boundary. The confining boundary is constructed out of identical magnetic links. Each link is a simple beam with an embedded permanent magnet at each end. The links are secured to one another through the magnetic attraction present when the links are stacked end-to-end (Fig. \ref{fig:concept}a). Once the links are secured to one another, the magnetic attachments act as hinges that are free to rotate within the x-y plane (resembling the flexible behavior of a bike chain). Such flexibility, allows the boundary to morph into various geometric configurations. All the embedded magnets in the meta-atoms and the boundary are oriented in the same direction (i.e., north pole facing upward). The repulsive forces between the disk's magnets and the magnetic boundary cause the disks to self-assemble into their lowest potential energy locations. The final assembly is a function of \textit{the number of magnetic particles} and \textit{the geometry of the confining magnetic boundary} (Fig. \ref{fig:concept}b). 


\begin{figure*}
	\includegraphics[scale=1.1]{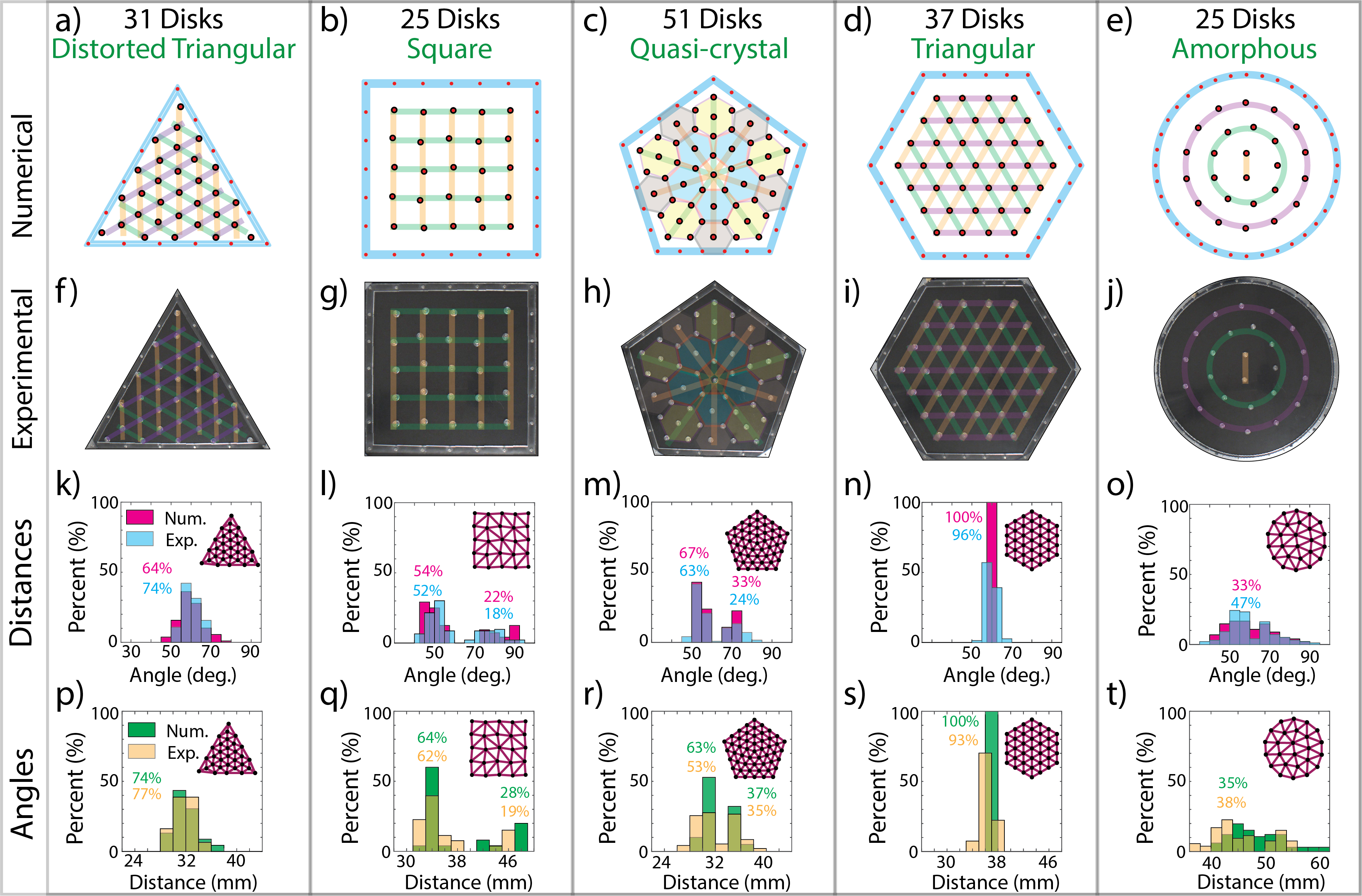}
	\caption{\label{fig:fixed_ass} \textbf{Self-assembly of different disk numbers and boundaries.} (a-e) Numerical and (f-j) experimental self assembly of (a, f) 31 disks within a triangle boundary, (b, g) 25 disks within a square boundary, (c, h) 51 disks within a pentagon boundary, (d, i) 37 disks within a hexagon boundary, and (e, j) 25 disks within a circle boundary. (k-t) Numerical (opaque) and experimental (transparent) histograms of the (k-o) distances and (p-t) angles between the self assembled disks. The Delaunay triangulations are shown in the insets.}
\end{figure*}

To study the self-assembly of meta-atoms within a confining boundary, we systematically vary both the number of magnetic particles and the number of edges of the magnetic boundary\cite{norouzi2021classification}. We consider boundaries with a number of edges, $n$ = 3-6 (i.e., triangle, square, pentagon, and hexagon) in addition to $n = \infty$ (i.e., circle). We simulate the self-assembly of the particles using the Newmark method\cite{press2007numerical} and observe the emerging pattern. All disks start at random positions within the boundary. Then, the position of each disk evolves over time, governed only by the repulsive magnetic forces acting on them. It is important to note that no external shaking of the boundary or the disks is required. Eventually, the system reaches its equilibrium state where a pattern emerges. Within each confining boundary shape, there exist multiple intriguing stable patterns with endless possibilities that emerge regardless of the initial positions of the disks. As an example, we show case one pattern per boundary-shape with a given number of magnetic disks (Fig. \ref{fig:fixed_ass}a-e). For a triangle boundary confining 31 disks, a distorted triangular lattice emerges  (Fig. \ref{fig:fixed_ass}a). Within the square boundary, 25 disks self assemble into a square lattice (Fig. \ref{fig:fixed_ass}b). Within the pentagon boundary, 51 disks emerge into a quasi-crystal pattern with five-fold rotational symmetry  (Fig. \ref{fig:fixed_ass}c). For 37 disks confined within a hexagon boundary, a triangular lattice emerges (Fig. \ref{fig:fixed_ass}d). Within the circle boundary, 25 disks self assemble into an amorphous pattern (Fig. \ref{fig:fixed_ass}e). 

To validate our numerical findings, we replicate the simulations experimentally. We start by placing each simulated boundary shape with the specified number of  randomly positioned disks on an air-bearing surface (air-hockey table). When the air-bearing is activated, the disks float on a thin layer of air which minimizes the friction between the disk's bottom surface and the air-bearing's surface allowing the magnetic repulsion forces to dominate the assembly dynamics. For all considered boundary shapes, we observe the emergence of identical patterns to the simulated ones  (Fig. \ref{fig:flex_assembly} f-j).  

To analyze the resultant patterns, we perform Delaunay triangulation on the  disk equilibrium positions within the five boundaries (Fig.  \ref{fig:fixed_ass}k-t). We calculate the histogram distributions of the distances (bin width 2 mm) and angles (bin width 5$^\circ$) of each lattice. Immediately, we note the stark difference in the distribution of distances and angles between the case of the hexagonal boundary (Fig. \ref{fig:fixed_ass}n,s) and the circular boundary (Fig. \ref{fig:fixed_ass}o,t). In the hexagonal boundary, for both the distances between disks (Fig. \ref{fig:fixed_ass}n) and the angles of the lattice (Fig. \ref{fig:fixed_ass}s), the distribution is very narrow, indicating uniformity throughout the lattice. In the circular boundary, the distribution for both the distances and the angles (Fig. \ref{fig:fixed_ass}t) between the disks (Fig. \ref{fig:fixed_ass}o) is very wide, indicating variability throughout the assembly. 

To quantify the emerging order, or the lack thereof, of the numerically simulated emergent lattices, we define a crystallinity index (C.I.) as the addition of the two highest histogram bin values to account for statistical variance. In the case of the hexagonal boundary, the addition of the two highest bins for both the angles of the lattice and distances between the disks, is 100\%, indicating high crystallinity within the lattice (Fig. \ref{fig:fixed_ass}n,s).  Conversely, in the case of the circular boundary, the C.I. for the angles of the lattice and distances between adjacent disks is 33.33\% and 34.85\%, respectively, indicating an amorphous assembly (i.e., no crystallinity) (Fig. \ref{fig:fixed_ass}o,t). In the case of the square boundary, we observe two bell shaped histograms for both the histogram of the lattice angles (Fig. \ref{fig:fixed_ass}q) and the histogram of the distances between the disks (Fig. \ref{fig:fixed_ass}l). This distribution results from the geometry of the isosceles-resembling triangles formed during Delaunay triangulation. The distribution aligns with the inherent characteristic of an isosceles triangle (i.e., two smaller angles and one larger angle, and two similar side lengths and one differing side length). We quantify this distribution by labeling the C.I. of each bell curve of the lattice angles and disk distances histograms. In the case of the pentagonal boundary, we observe a similar trend of two distinct bell curves present in each histogram (Fig. \ref{fig:fixed_ass}m,r); however, this distribution is due to the formation of a quasi-crystalline lattice, instead of isosceles triangles from the Delaunay triangulation. We experimentally validate the numerical histogram distributions, by performing a Delaunay triangulation of the experimental self assemblies of the lattices, and observe the same trends (transparent bars in Figure  \ref{fig:fixed_ass}k-t).


\begin{figure}[!ht]
	\includegraphics[scale=.97]{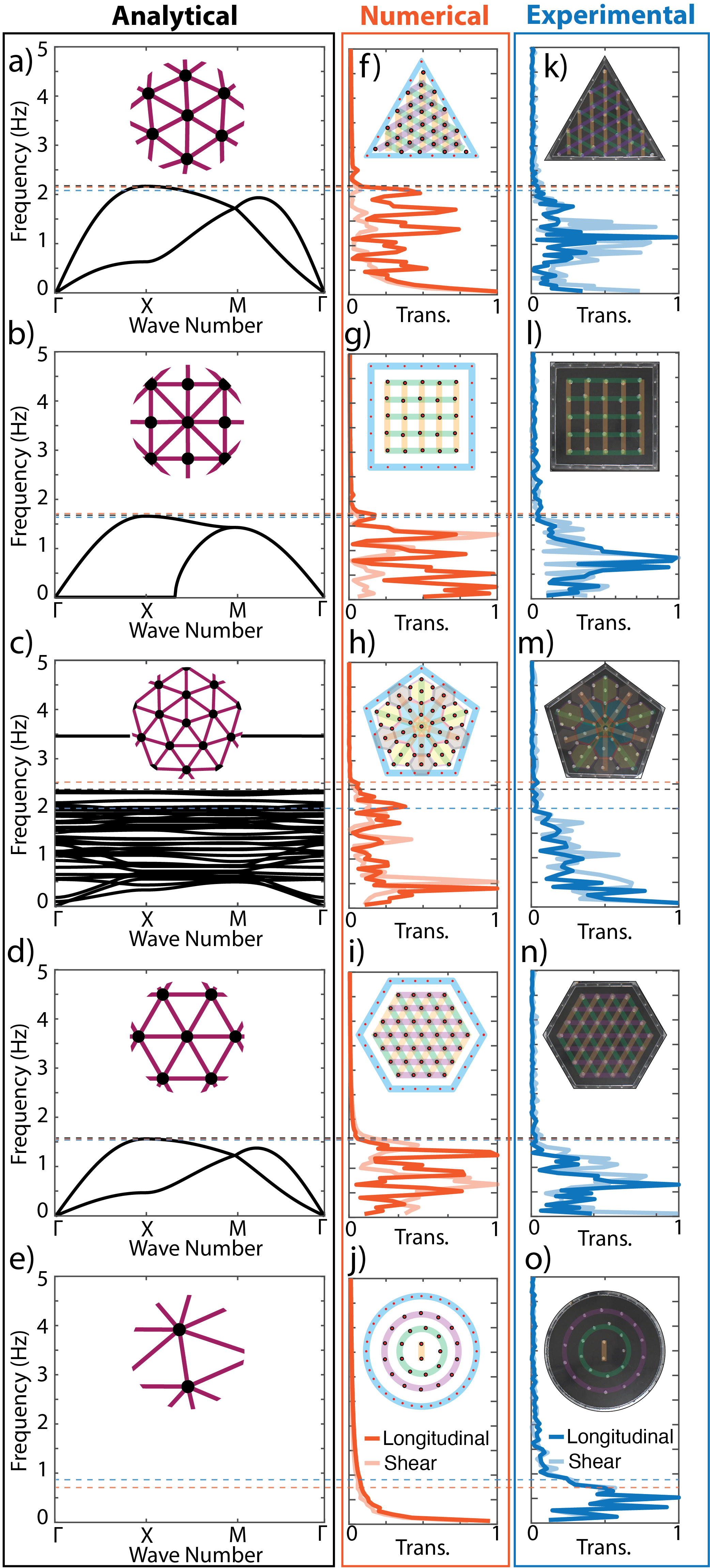}
	\caption{\label{fig:fixed boundary transmission} \textbf{Wave transmission within the self-assembled patterns.}  (a-e) Dispersion curves of the emergent unit cells within each assembly for (a) triangular, (b) square, (c) quasi-crystal, and (d) triangular lattice. (f-j) Numerical wave transmission within the (f) triangle, (g) square, (h) pentagon, (i) hexagon, and (j) circle boundary. (k-o) Experimental wave transmission within the (f) triangle, (g) square, (h) pentagon, (i) hexagon, and (j) circle boundary.}
\end{figure}

To understand the dynamical behavior of the emerging patterns, we consider their wave transmission characteristics. Each pattern, aside from the quasi-crystal assembly within the pentagon boundary and the amorphous assembly within the circle boundary, possesses translational symmetry. The symmetry allows for the conventional use of Bloch analysis to determine the transmission characteristics through each lattice. The resulting dispersion curves represent a correlation between frequency and wave number, revealing the expected transmission regions within the frequency spectrum (Fig. \ref{fig:fixed boundary transmission}a, b\&d). 
To obtain the dispersion curves of the quasi-crystal lattice, we implement the supercell method\cite{chen2015band}, where we derive the equations of motion of an artificial unit cell consisting of 19 central disks (i.e., masses) connected by springs and dampers (see supplemental material) and assume infinite repetition of this unit cell in space (Fig. \ref{fig:fixed boundary transmission}c). 
For the amorphous lattice within the circular boundary, there exists no dispersion curves as there is no translational nor rotational symmetry (Fig. \ref{fig:fixed boundary transmission}e).

\begin{figure*}
\includegraphics[scale=0.93]{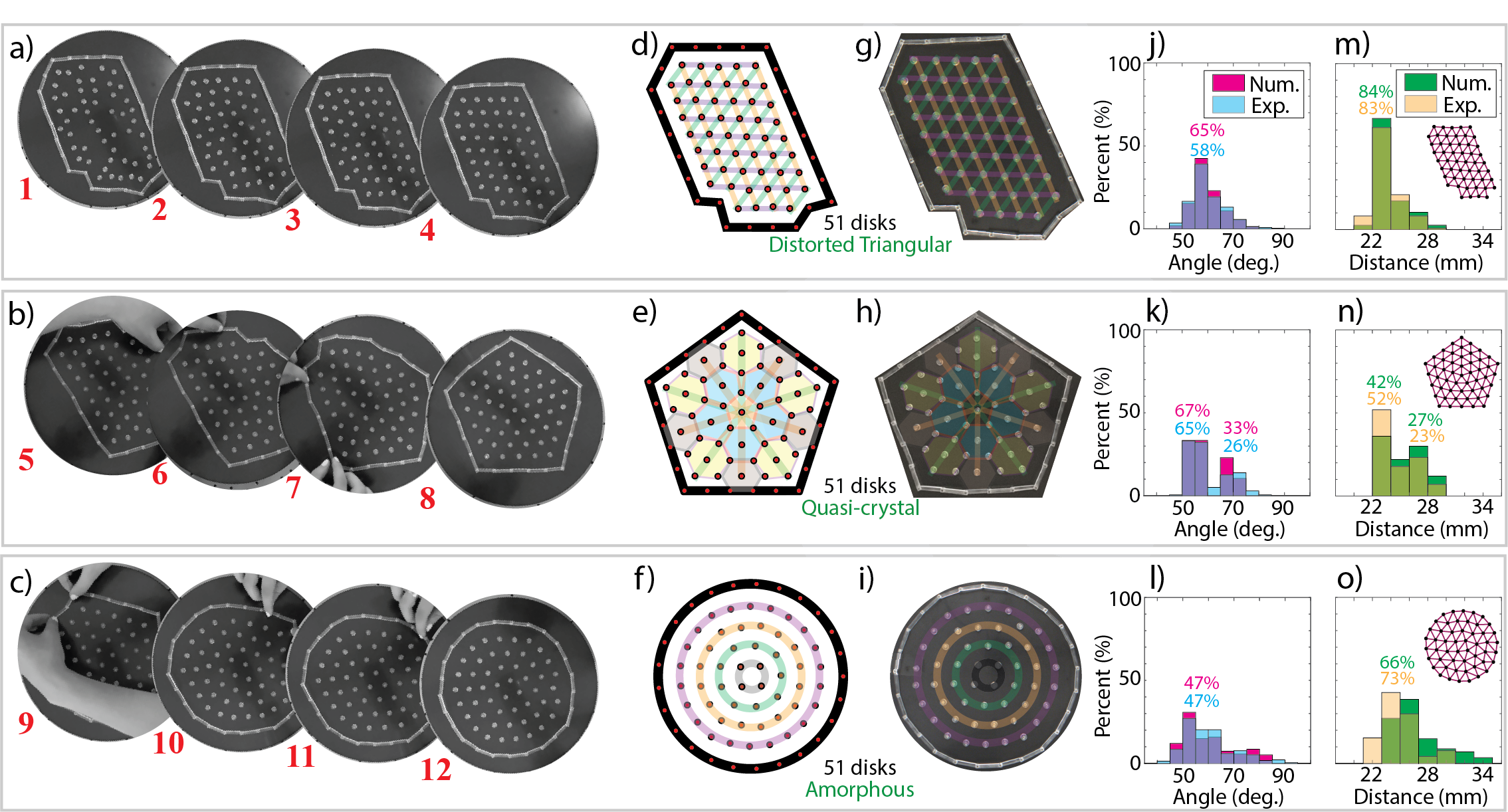}
\caption{\label{fig:flex_assembly} \textbf{Re-programmable self-assembly.}  (a) Experimental self-assembly of 51 disks within  an abstract boundary shape (1-4). (b) Evolution of the magnetic boundary from an abstract shape to a pentagon (5-8). (c) Evolution of the magnetic boundary from a pentagon shape to a circle (9-12).  (d-f) Numerical and (g-i) experimental self-assembly of the 51 disks within the (d, g) abstract, (e, h) pentagon, and (f, i) circle boundary. (j-o) Numerical (opaque) and experimental (transparent) histograms of the (j-l) angles and (m-o) distances between the self assembled disks. The Delaunay triangulations are shown in the insets.}
\end{figure*}

To numerically validate the analytical dispersion curves, we simulate the frequency response function (FRF) of each lattice using the Verlet method. We excite each lattice with a chirp signal in the direction of a specified lattice vector (see supplemental material), and calculate the fast Fourier  transform (FFT) of the displacement of a disk along the direction of the lattice vector to obtain the numerical wave transmission through each lattice (Fig. \ref{fig:fixed boundary transmission}f-j). The numerical transmission spectra for each lattice match very well with the analytical dispersion curves. 
To experimentally verify the numerical and analytical transmission spectra, we excite each lattice with a chirp signal and track the motion of the disks in time using an overhead camera. We post-process the images and perform an FFT on the temporal signal of one the freely floating disks to obtain the experimentally observed transmission within each lattice (Fig. \ref{fig:fixed boundary transmission}k-o). The experimental transmission for each lattice matches well with the analytical and numerical transmission, with the tendency of having a slightly lower upper transmission cutoff, except in the case of the amorphous lattice within the circular boundary. 


All realized patterns are remarkable in their own accord as they span varying levels of order and symmetry, however, each pattern contains different number of meta-atoms, boundary magnets,  perimeter length (i.e., magnetic links), and area. This translates to significant change in the required conditions of the assembly to alternate from one phase of matter to another. To overcome this limitation, and make it rather simple to shift between phases, we find a single number of meta-atoms that can assemble into different states (crystalline, quasi-crystal, and amorphous) using the same perimeter length and number of boundary magnets. We start by considering an abstract arrangement of the boundary links with 51 confined meta-atoms resting at random positions (Fig. \ref{fig:flex_assembly}-a1). Once the air bearing is activated, friction is minimized and the intrinsic magnetic repulse force drive the meta-atoms to rearrange themselves (Fig. \ref{fig:flex_assembly}a(2-4)), eventually settling in an arrangement that resembles a distorted triangular lattice. With 51 disks within the abstract boundary, we observe the emergence of a distorted triangular lattice. To ensure the stability of the emerging assembly, we repeat the experiment numerous times using the same boundary shape, but with different random initial positions of the disks. Regardless of the initial disk positions, the final emerging pattern is always a distorted triangular lattice (Fig. \ref{fig:flex_assembly} g). Furthermore, to verify our findings, we simulate the self-assembly numerically using the Verlet method \cite{press2007numerical}. Similar to the experiments, we start the disks at random positions and allow their positions to evolve over time, governed by the repulsive magnetic forces, until the system reaches its equilibrium state (Fig. \ref{fig:flex_assembly} d). Once more, regardless of the initial positions  of the disks, they always self-assemble in the same distorted triangular lattice. Without deactivating the air-bearing, removing or adding disks, or modifying the boundary linkages, we reshape the magnetic boundary into a pentagonal shape in real-time and continue to monitor the evolution of the assembly (Fig. \ref{fig:flex_assembly} b).  The disks self-assemble into a quasi-crystalline pattern (Fig. \ref{fig:flex_assembly} h), with no clear translational symmetry. The final assembly features a pentagon pattern at the center, surrounded by hexagons from all sides. Moreover, we also simulate the process numerically and obtain the same final quasi-crystalline pattern (Fig. \ref{fig:flex_assembly} e). Finally, we reshape the boundary further to resemble a circular shape, while keeping the air-bearing on, and continue to record the evolution  of the assembly (Fig. \ref{fig:flex_assembly} c).  The final assembly, similar to that obtained in a parallel numerical simulation, is amorphous with no clear symmetry (Fig. \ref{fig:flex_assembly} f,i). 


\begin{figure}
\begin{center}
\includegraphics[scale=0.95]{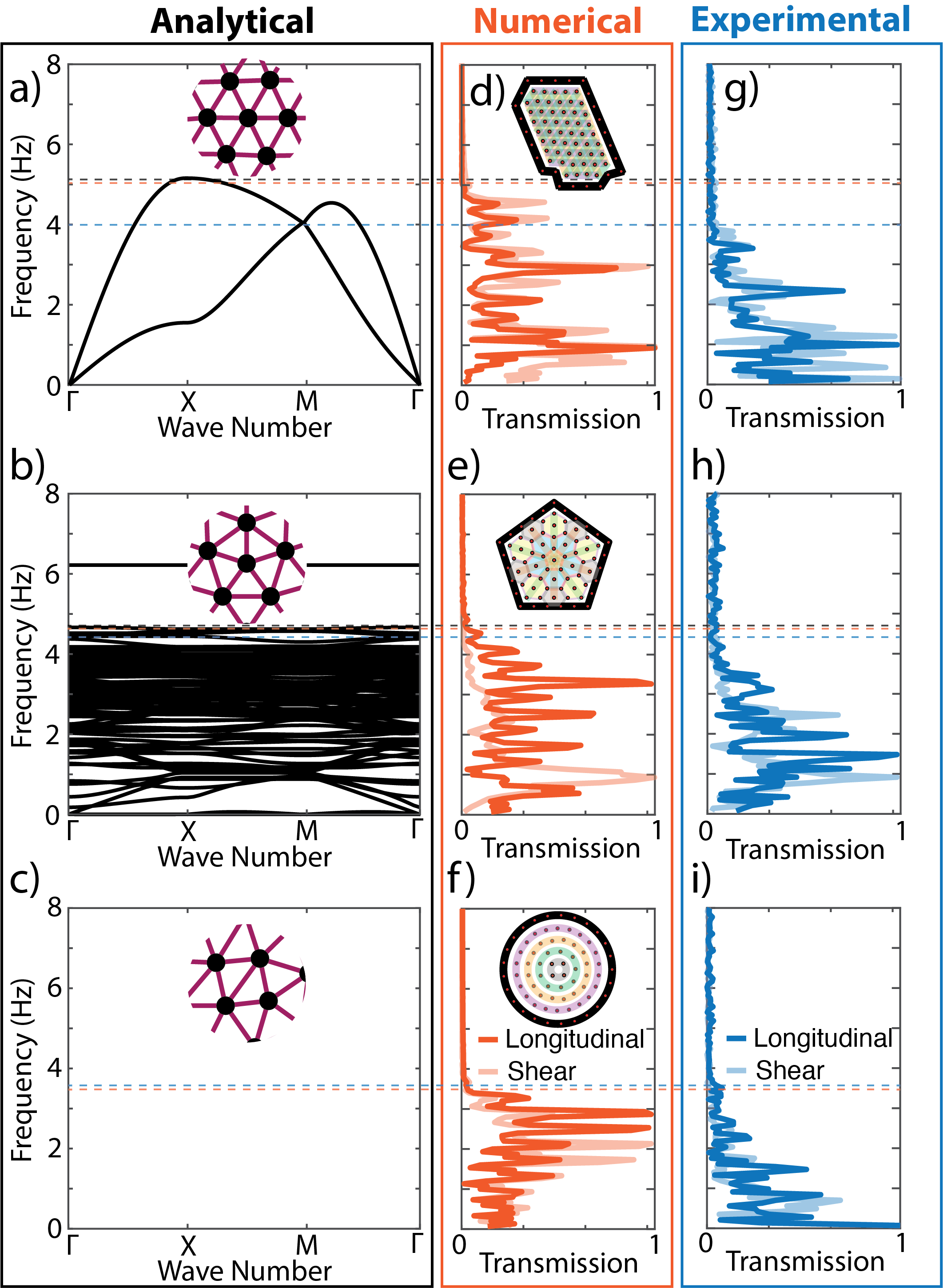}
\caption{\label{fig:flexible boundary transmission} \textbf{Re-programmable wave transmission.}  (a-c) Dispersion curves of the emergent unit cells within each assembly for (a) triangular and (b) quasi-crystal lattice. (d-f) Numerical wave transmission within the (d) triangle, (e) pentagon, and (f) circle boundary. (g-i) Experimental wave transmission within the (g) triangle, (h) pentagon, and (i) circle boundary.}
\end{center}
\end{figure}

To further quantify the emerging patterns and estimate the intrinsic order (crystallinity) of the lattices within the flexible boundary, we use Delaunay triangulation (Fig. \ref{fig:flex_assembly} d-i). The distributions of lattice angles and distances between the fixed number of the meta-atoms and the morphing boundary closely resemble those of the previously obtained assemblies with similar features. The distorted triangular lattice within the morphing abstract boundary (Fig. \ref{fig:flex_assembly} j, m) and the triangular boundary (Fig. \ref{fig:fixed_ass} k, p) both show unimodal distributions, highlighting their similar lattice structures. In the case of the pentagonal boundary, the distribution of the lattice angles and distances are also two distinct bell curves, similar to those reported in Figure \ref{fig:fixed_ass}m,r. In the case of the circular boundary, the distributions of the lattice angles and disk distances are single bell curves, that have much larger distributions in comparison to the abstract and pentagon boundary shapes (Fig. \ref{fig:flex_assembly} l,o). We utilize the Delaunay triangulation on the experimental self assemblies to quantify and experimentally verify the lattice formations. We determine the crystallinity index of the distances between adjacent disks and angles of the lattices to be very close to those of the numerical self assemblies (Fig. \ref{fig:flex_assembly} j-o).

To understand the dynamical behavior of the morphable patterns, we consider their wave transmission characteristics. For the abstract boundary, we can define the unit cell of the lattice, assume infinite periodicity, and analytically determine the dispersion curves and transmission using Bloch's theorem \cite{bloch1929quantenmechanik} (Fig. \ref{fig:flexible boundary transmission}a). The lattice present within the pentagon boundary has no definitive unit cell and therefore no translational symmetry, preventing the use of conventional Bloch analysis; however, the same Supercell Method employed earlier can model the transmission through a quasi-crystalline lattice remarkably well \cite{chen2015band} (Fig. \ref{fig:flexible boundary transmission}b). The amorphous lattice present within the circular boundary (Fig. \ref{fig:flexible boundary transmission}c) has no translational symmetry nor rotational symmetry, rendering the analytical modeling of transmission through the lattice to be implausible. To numerically validate the analytical dispersion curves, we simulate the frequency response function (FRF) of each assembly using the Verlet method. We excite each lattice with a chirp signal in the direction of a specified lattice vector (see supplemental material) and take a fast Fourier transform (FFT) of the displacement (in the direction of excitation) of a specified disk within the lattice vector (Fig. \ref{fig:flexible boundary transmission}d-f). The numerical transmission within the abstract and pentagon boundary match relatively well with the analytical dispersion curves. We validate the transmission ranges of the self-assembled patterns experimentally. The experimental transmission matches relatively well with the analytical and numerical transmission, with the tendency to have a slightly lower upper transmission frequency cutoff. 



In summary, we have analytically, numerically, and experimentally demonstrated the dynamic tunability (i.e., reprogramming) of  the self-assembly of magnetic particles. The resulting patterns span a varying level of order and symmetry, with our assemblies showing crystalline, quasi-crystalline, and amorphous patterns. Furthermore, we demonstrate the transition between all three phases with the same number of particles, boundary magnets, links and perimeter. In the instances of translational (i.e., distorted triangular) or rotational symmetry (i.e., quasi-crystal), we analytically modeled and numerically and experimentally verified the transmission ranges through each lattice. Our platform can be utilized to manipulate ultra-low frequency waves, within a relatively small space. The inherent nonlinear potentials within each lattice can be harnessed to demonstrate phenomena with no linear parallel such as amplitude dependent response, bifurcation, chaos and solitons.\cite{spadoni2010generation, mehrem2017nonlinear, porter2015granular, gendelman2018introduction, kim2019wave, ramakrishnan2020transition, amendola2018tuning, patil2022review} In addition, the self-assembly of the different lattices can be key in creating re-programmable materials with exceptional properties.\cite{culha2020statistical,samak2024evidence,samak2024direct,stenseng2024bi}




\bibliography{sample}

\appendix

\beginsupplement
\newpage
\onecolumngrid

\section{Experimental methods}
To fabricate the flexible boundary, we cut each link from acrylic glass using a laser cutter (Full-Spectrum 24 Pro Series), embed the 3 $mm$ $\times$ 3 $mm$ cylindrical permanent magnets 29 $mm$ apart from one another, and assemble the confining boundary shape atop an air-bearing table (New way S1030002). To fabricate the magnetic particles, we laser cut 51 acrylic glass disks, each having a diameter of 8 $mm$, concentrically embed a 3 $mm$ $\times$ 3 $mm$ cylindrical permanent magnet, and adhere a glass slide to the underside of the magnet-embedded disk to further reduce friction.  To excite each lattice, we use a mechanical shaker (type 4810) and function generator (Keysight Technollgies 33512B). We use an overhead camera (Blackfly-S USB3) to capture the motion of the disks in time and post-process the images using the digital image correlation software (DICe) to obtain the displacement of each disk in time.  

\section{Dispersion analysis for ordered lattices}

For each of the considered systems, the eigenvalue problem for the unit cell composed of a single floating disk can be written as: 
\begin{equation}
[-\omega^2\textbf{M}+i\omega\textbf{C}(\boldsymbol\kappa)+\textbf{K}(\boldsymbol\kappa)] \boldsymbol{\phi} = 0,
\label{eqn:eigen}
\end{equation}
where $\omega$ is the frequency, $\textbf{M} = \begin{bmatrix}
m & 0\\
0 & m
\end{bmatrix}$ is the mass matrix where $m=0.305~g$ is the mass of a magnet-embedded disk, $i=\sqrt{-1}$, $\textbf{C}(\boldsymbol\kappa)$ is the damping matrix as a function of wave number $\boldsymbol\kappa$, $\textbf{K}(\boldsymbol\kappa)$ is the stiffness matrix as a function of wave number, and $\boldsymbol\phi = [u~v]^T$ is the Bloch displacement vector in the $x$ and $y$ directions. The stiffness and damping matrices are written as
\begin{equation}
\mathbf{K}(\boldsymbol\kappa)=2\displaystyle \sum_{i=1}^{p}\left \{  {f_{,d}(d_i)\boldsymbol{e_i}\otimes \boldsymbol{e_i}[cos(\boldsymbol\kappa\cdot \boldsymbol{R_i})-1]}\right \}+2\displaystyle \sum_{i=1}^{p}\left \{ {\frac{f(d_i)}{d_i}(\boldsymbol{I}-\boldsymbol{e_i}\otimes \boldsymbol{e_i})[cos(\boldsymbol\kappa\cdot \boldsymbol{R_i})-1]} \right \}
\label{eqn:2D stiff}
\end{equation} 
\begin{equation}
\mathbf{C}(\boldsymbol\kappa)=2\displaystyle \sum_{i=1}^{p} {c_{lattice}\boldsymbol{e_i}\otimes \boldsymbol{e_i}[cos(\boldsymbol\kappa\cdot \boldsymbol{R_i})-1]}
\label{eqn:2D damp}
\end{equation} 
respectively, where $p$ is the number of unit vectors that defines the unit cell, $d_i$ is the mode value of the distances between the disks in the lattice in the $i^{th}$ direction, $\boldsymbol{e_i}$ are the unit vectors, $\otimes$ is the dyadic product, $\boldsymbol{R_i}=L\boldsymbol{e_i}$ are the lattice vectors where $L$ is the distance between disks along the same lattice vector, and $c_{lattice}$ is the experimentally determined damping value of the lattice. 
The stiffness matrix in equation \ref{eqn:2D stiff} takes into account the repulsive forces between the magnetic particles following an inverse power law in the form $f(d) = A d^{\gamma}$ and $f_{,d}(d)$ as its first derivative. For a dipole-dipole interaction \cite{watkins2020demultiplexing}, $A = 3\mu B^2/4\pi$, where $\mu$ is the permeability of air and $B$ is the magnetic moment. We perform a Bloch state space transformation \cite{deymier2013acoustic} to formulate the transformed eigenvalue problem:
 \begin{equation}
[-\omega^2\textbf{$\overline{\textbf{M}}$}(\boldsymbol\kappa)+\textbf{$\overline{\textbf{K}}$}(\boldsymbol\kappa)] \boldsymbol{\phi} = 0,
\label{eqn:eigen transformed}
\end{equation}

where $\boldsymbol {\overline{M}}(\boldsymbol\kappa) = \begin{bmatrix}
\textbf{0} & \textbf{M}\\
\textbf{M} & \textbf{C}(\boldsymbol\kappa)
\end{bmatrix}$
and $\boldsymbol {\overline{K}}(\boldsymbol\kappa) = \begin{bmatrix}
-\textbf{M} & \textbf{0}\\
\textbf{0} & \textbf{K}(\boldsymbol\kappa)
\end{bmatrix}$
are the transformed mass and stiffness matrices, respectively.

To obtain the frequency-wave number relationship of the appropriate lattices, we assemble the stiffness (Eqn. \ref{eqn:2D stiff}) and damping matrices (Eqn. \ref{eqn:2D damp}) where the unit vectors for the lattice within the triangle and hexagon boundary are $\boldsymbol{e_1}=[1, 0]^T$, $\boldsymbol{e_2}=[0.5, \sqrt{3}/2]^T$, and $\boldsymbol{e_3}=[-0.5, \sqrt{3}/2]^T$, within the square boundary are $\boldsymbol{e_1}=[1, 0]^T$, $\boldsymbol{e_2}=[0, 1]^T$, and formulate and solve the transformed eigenvalue problem (Eqn. \ref{eqn:eigen transformed}) to obtain the dispersion curves using a value of $L=31~mm$ for the lattice within the triangle boundary, $L=37~mm$ for the lattice within the hexagon boundary, or $L=34~mm$ for the lattice within the square boundary (Fig. \ref{fig:fixed boundary transmission}a, d, b). By solving equation \ref{eqn:eigen transformed}, using $p=3$ lattice vectors, the unit vectors are $\boldsymbol{e_1}=[1, 0]^T$, $\boldsymbol{e_2}=[0.5, \sqrt{3}/2]^T$, and $\boldsymbol{e_3}=[-0.5, \sqrt{3}/2]^T$, $L=23~mm$, and using the experimentally determined damping value $c_{abstract}=5.35\times10^{-4}$ kg/s, we obtain the dispersion curves of the distorted triangular lattice (Fig. \ref{fig:flexible boundary transmission}a).

\section{Pentagon lattice analytical modeling}

\begin{figure*}[h]
\includegraphics[scale=1]{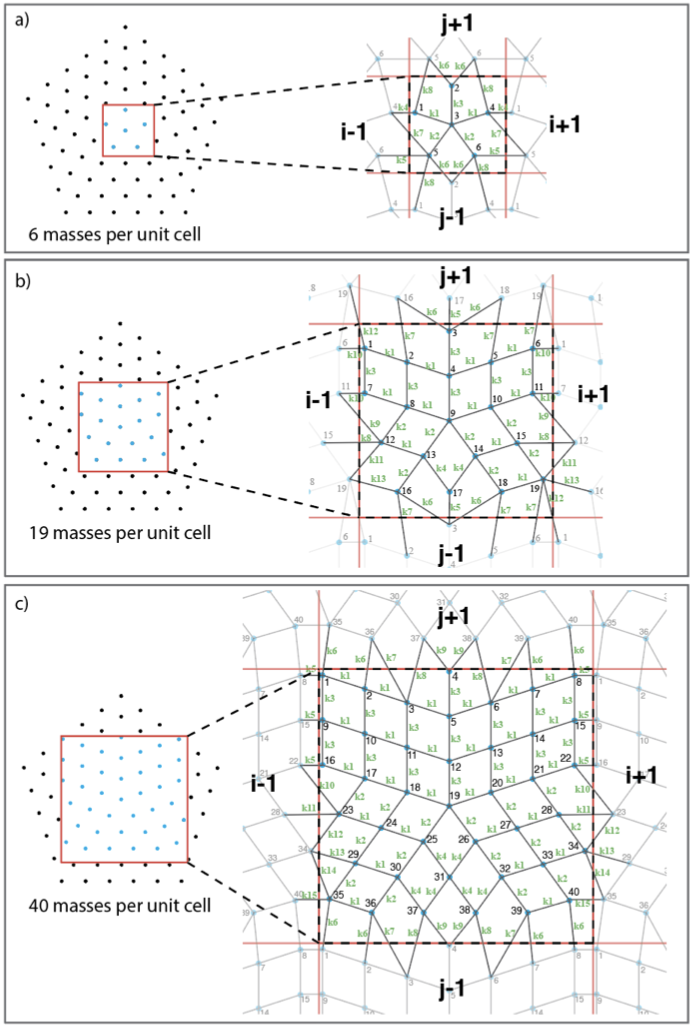}
\caption{\label{fig:pent unit cells} \textbf{Unit cell couplings.} Defined unit cells (left) and the mass-spring couplings (right) of unit cells with a) 6, b) 19, and c) 40 masses.}
\end{figure*}

To obtain the analytical transmission through the quasi-crystalline lattices (Fig. \ref{fig:fixed boundary transmission}c and \ref{fig:flexible boundary transmission}b), we implement the supercell method \cite{chen2015band}.  We start by numerically modeling the self-assembly of a lattice with 76 disks confined within a pentagon-shaped boundary and ensure the self-assembly of these disks match the locations of the disks seen within the flexible boundary pentagon lattice. Then, we define three unit cells of three different sizes resulting in three different number of disks per unit cell: (1) 6 disks, (2) 19 disks, and (3) 40 disks (Fig. \ref{fig:pent unit cells} (a-c, left)). We define couplings between the disks such that the unit cell has translational symmetry when repeated in the x-direction ($i$, $i+1$, $i-1$) and y-direction ($j$, $j+1$, $j-1$) (Fig. \ref{fig:pent unit cells} (a-c, right)). We calculate the static repulsive forces between each pair of disks using the inverse power law $f(d) = A d^{\gamma}$ and $f_{,d}(d)$ as its first derivative where $A = 9.617\times10^{-11}$, $\gamma=-4$, where $d$ is determined through the numerical simulations for each individual disk coupling. 

\begin{figure*}
\includegraphics[scale=0.45]{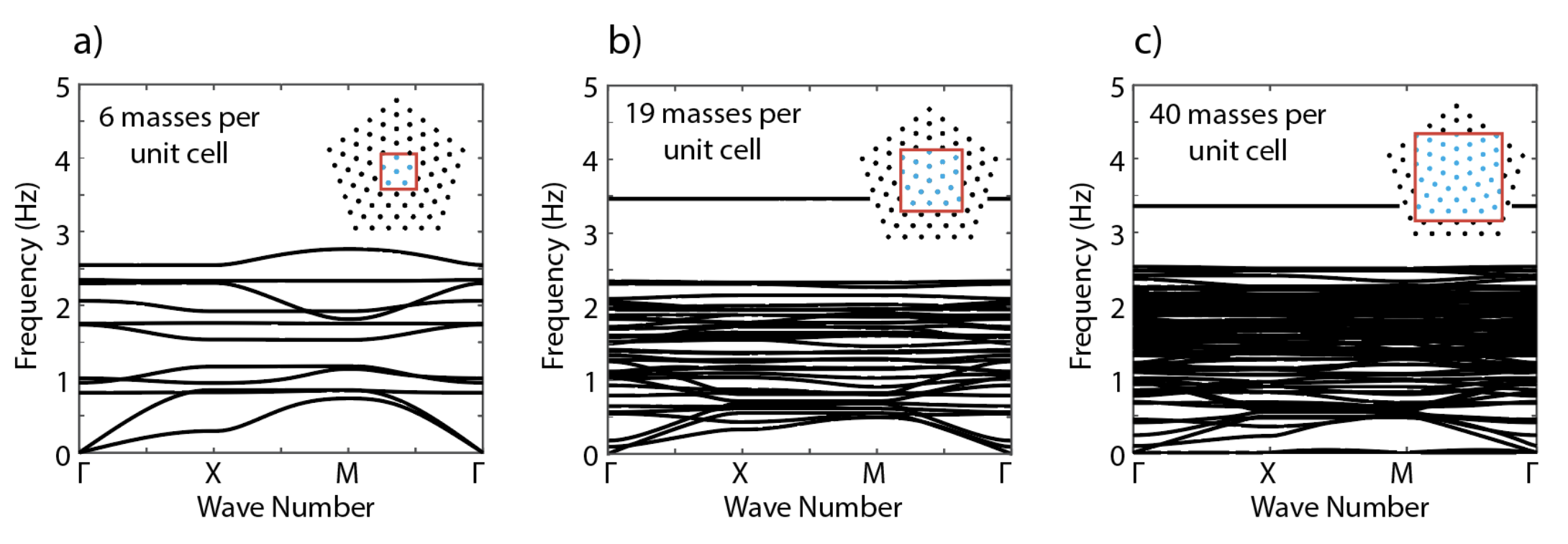}
\caption{\label{fig:pent disp fixed boundary} \textbf{Pentagon boundary dispersion from figure 3.} Generated dispersion curves for three distinct unit cells defined within the pentagon lattice consisting of 76 disks. Dispersion curves generated for unit cell with a) 6, b) 19, and c) 40 masses.}
\end{figure*}
\begin{figure*}
\includegraphics[scale=0.45]{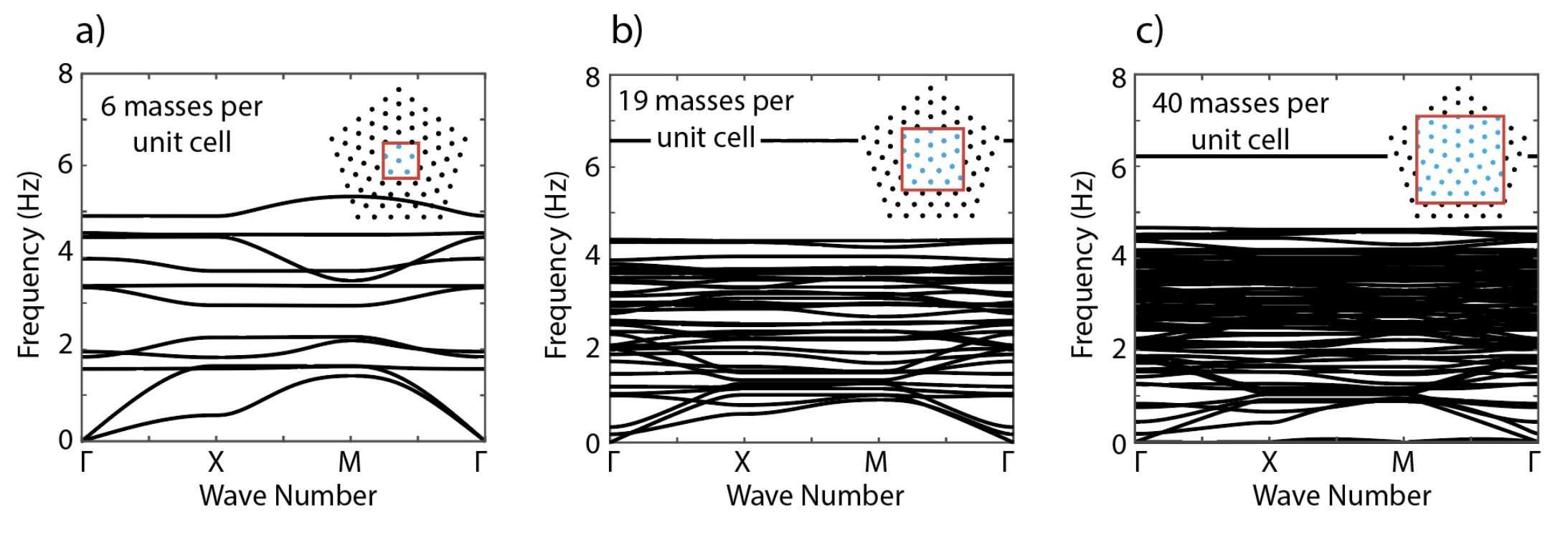}
\caption{\label{fig:pent disp flex boundary} \textbf{Pentagon boundary dispersion from figure 5.} Generated dispersion curves for three distinct unit cells defined within the flexible pentagon lattice consisting of 76 disks. Dispersion curves generated for unit cell with a) 6, b) 19, and c) 40 masses.}
\end{figure*}

We utilize the experimentally determined damping value for the lattice in figure 3, $c=3.114\times10^{-4}$ kg/s and figure 5, $c=5.73\times10^{-5}$ kg/s, and derive the equations of motion for each unit cell. From the equations of motion, we formulate the mass, stiffness, and damping matrices, assume infinite repetition of each unit cell (in order to use Bloch analysis), and assemble and solve the transformed eigenvalue problem (Eqn. \ref{eqn:eigen transformed}) to generate the three sets of dispersion curves of the quasi-crystalline lattice in both Figure 3 and 5 with three different unit cells (Figs. \ref{fig:pent disp fixed boundary} and \ref{fig:pent disp flex boundary}). In the case of 6 masses per unit cell, we observe stop bands that does not match with either the numerically nor the experimentally obtained transmissions (Fig. \ref{fig:pent disp flex boundary}a). As the number of masses included in the unit cell increases, the stop bands (regions of attenuation) shrink, but the overall span of the transmission region remain relatively similar.


\begin{figure*}
\includegraphics[scale=0.5]{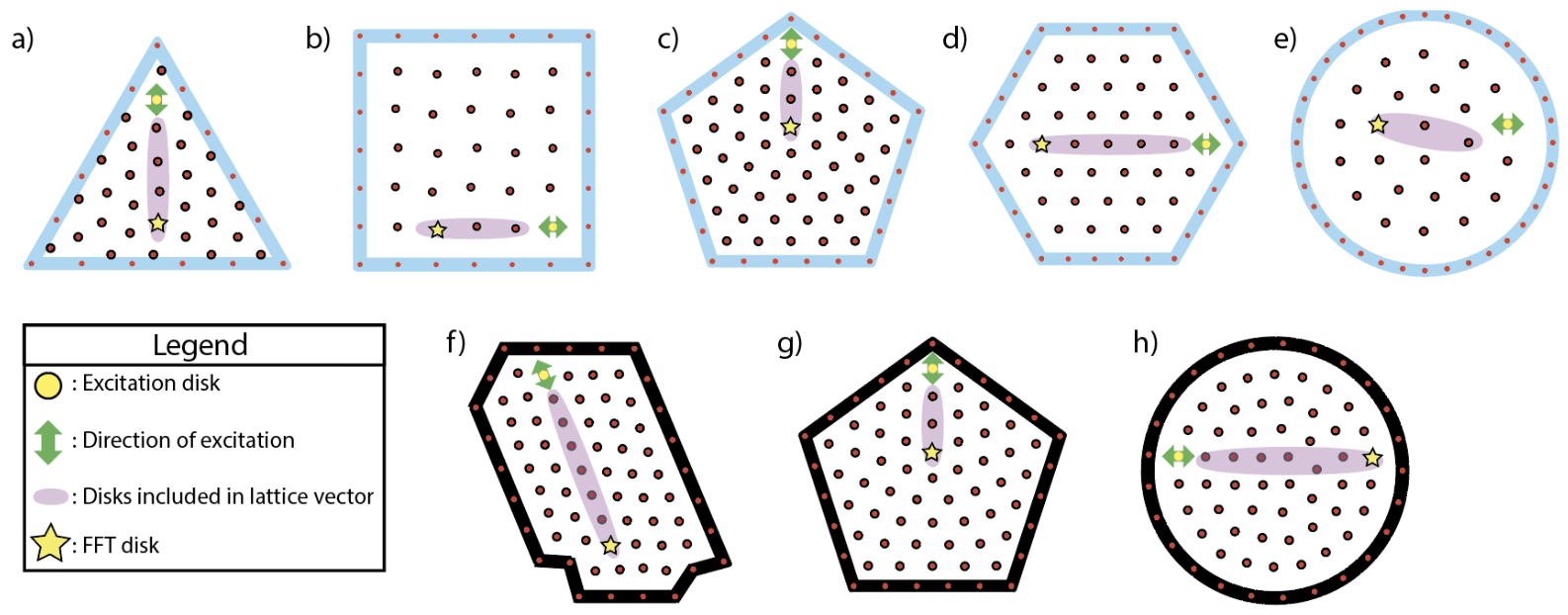}
\caption{\label{fig:lattice vectors} \textbf{Defined lattice vectors.} Numerically simulated self-assembly of lattices with fixed (blue) and flexible (black) boundaries confined to the a) triangle, b) square, c) pentagon, d) hexagon, e) circle, f) flexible abstract, g) flexible pentagon, and h) flexible circle boundaries. The excitation disk, direction of excitation, disks included in lattice vector, and disk the FFT was performed on are shown within each lattice and labeled by the legend.}
\end{figure*}

\section{Wave transmission analysis}

To obtain the numerically simulated and experimentally verified transmission data (FFTs) (Fig. \ref{fig:flexible boundary transmission}(g-i)(ii, iii), \ref{fig:fixed boundary transmission}(k-o)(ii, iii)), excited each lattice with a single excitation disk in the direction of a defined lattice vector, and performed an FFT on a single disk within the same lattice vector (Fig. \ref{fig:lattice vectors}). The lattice within each boundary shape has its own unique lattice vector that remains consistent throughout the numerical simulations and experimental testing. The FFT of the FFT disk is performed on the displacement of the disk parallel to the direction of excitation.

\newpage

\end{document}